\documentclass[reprint,aps,amsmath,amssymb]{revtex4}
\usepackage[]{graphicx}
\usepackage{subfigure}
\usepackage{times}
\usepackage{color}

\newcommand{\comment}[1]{}

\newcommand{\be}{\begin{equation}}
\newcommand{\ee}{\end{equation}}
\newcommand{\bea}{\begin{eqnarray}}
\newcommand{\eea}{\end{eqnarray}}
\newcommand{\edu}{\end{document}}

\begin{document}

\title{Dielectric constant of
 monolayer transition metal
dichalcogenides across excitonic resonances}

\author{A. Thilagam}
\email[]{thilaphys@gmail.com}

\affiliation{Information Technology, Engineering and Environment,\\ 
University of South Australia, Australia
 5095.}
\begin{abstract}
We analyze the dielectric-function spectra of low dimensional transition metal
dichalcogenides  (TMDCs)  using a fully analytical  model  of the 
complex dielectric function that is applicable
in fractional  dimensional space. We extract the dimensionalities 
of the $A$ and $B$ excitons  as well as their 
Lorentzian broadening widths by fitting the model to experimental data
 in  the spectral range of photon energies (1.5 - 3 eV).
Our results show the significant contribution of the lowest ground exciton state to
 the dielectric properties
of   exemplary monolayer materials (MoS$_2$, MoSe$_2$ and WSe$_2$).
The   exciton dimensionality 
 parametrizes the processes that underlie  confinement and many-body Coulomb effects
as well as substrate screening effects, which 
simplifies the analysis 
 of electro-optical properties in low dimensional systems.
This study  highlights the potential of  theoretical models as  valuable
tools for interpreting the optical spectrum and in seeking an understanding
of the correlated dynamics between the $A$ and $B$ excitons
on the dielectric function of  TMDCs. 
\end{abstract}

\maketitle


\section{Introduction}
The lattice dynamics\cite{molina2011phonons,zhao2013lattice,cai2014lattice,zheng2017phonon}
and dielectric properties \cite{park2018direct,
klots2018controlled,
li2014measurement,kumar2012tunable} of low dimensional transition metal dichalcogenides, MX$_2$
(M = Mo, W, Nb and X = S, Se)  are currently investigated  with great interest for both theoretical
studies
\cite{soh2018optical,
trushin2016optical,trolle2017model,berkelbach2015bright,kaasbjerg2012phonon,zhou2015berry,mukherjee2015complex,stroucken2015optically,trushin2016optical,chakraborty2013layer,steinhoff2016nonequilibrium,kolobov2016excitons,chow2017phonon},  and high-performance  device applications \cite{woodward2015wideband,ou2014ion,pospischil2014solar,radisavljevic2011integrated,li2016charge,fan2014mos2,
perebeinos2015metal,beck2000,tsai2013few,wi2014enhancement,xu2015one,
bertolazzi2013nonvolatile,ji2013epitaxial,yu2016evaluation,jariwala2017van,
park2016mos2,fengtunable,woo2017low}. Excitons are confined  strongly in low dimensional transition metal dichalcogenides (TMDCs) 
and display notable spectral features with desirable photoluminescence properties
 \cite{splendiani2010emerging,plechinger2012low,ji2013epitaxial,zhu2016strongly,eda2013two,bergh,molinaspin,mai2013many,gao2016localized,ghatak2011nature}. In Molybdenum disulfide (MoS$_2$), a
 well known member of the TMDCs, there exist two pronounced peaks which are linked to the direct gap $A$ and $B$ excitons. These peaks arise due to the   vertical transitions at the $K$ point from a spin-orbit split
valence band to a doubly degenerate conduction band with decrease in the number of lattice layers  \cite{makatom,rama,
chei12,sim2013exciton,komsa2012effects,qiu2013}. Ultrafast optical pump-probe spectroscopic measurements
display enhanced transient absorption blue-shifts for the  $A$ and $B$ excitons in the monolayer MoS$_2$  due to  repulsive inter-excitonic interactions,  with   non-trivial linewidth broadening effects \cite{sim2013exciton}.

The dielectric constant is an fundamental  quantity that underpins 
experimental observables such as the refractive index and absorption coefficient \cite{wemple1971behavior,
hopfield1958theory,takagahara1993effects,schmitt1989linear,yoffe2001semiconductor,tanguy1997analytical}
and provides valuable guidelines for the    fabrication of  optoelectronic and photonic devices.
Due to reduced contributions from ionic and surface  polarizabilities associated
with one or cluster of atoms, the dielectric constant decreases with increase in the frequency of the electric
field  \cite{hopfield1958theory,wemple1971behavior}.  Changes in the lattice structure that arise
from   frequency induced vibrations  also contribute to an overall decrease of the  polarization of the material.
It is well known that the decreased dielectric screening  and enhanced electron-electron correlation forces  
give rise to the   high  exciton binding energies noted in TMDCs
 \cite{makatom,bergh,qiu2013,chei12,ross2013,jones2013optical,rama,thiljap}.
Excitons are shown to dominate the dielectric properties of ultra-thin MoS$_2$ of less than 5-7 layers \cite{yu2015exciton}, with the dielectric function displaying an anomalous dependence  on the layer number. Currently,
there is lack of knowledge of the  effect of the 
correlated dynamics between the $A$ and $B$ excitons \cite{sim2013exciton}
on the dielectric function of  TMDCs. 

 A comprehensive  understanding of the role of excitons in the vicinity of the
 optical  region  provides useful guidelines in exploiting  the  dielectric properties 
of monolayer transition metal dichalcogenides for  novel applications such as
solar cells \cite{pospischil2014solar,tsai2014monolayer,tsuboi2015enhanced,liu2016fabrication,thilsolar2016},
 single-layer  transistors \cite{jariwala2013band,zhang2016highly} and light-emitting 
diodes \cite{woodhead2016light,jeon2015low,clark2016single}.
 These reasons  form the main motivation for this study where we employ
the  fractional dimension space approach (FDSA) \cite{still,he,matos99,christo93,mathieu,lefeb,oh1999geometric,reyes2000excitons,thil97stark,thilagam1999dimensionality,
lohe2004weyl} to examine the dielectric properties of the monolayer MoS$_2$
and related common TMDCs.  The FDSA maps an  
 anisotropic quantum quasi-particle in real space
to an isotropic environment parameterized by a single quantity $d$ ($1 \le d \le 4$) \cite{still,he,matos99,christo93,mathieu,lefeb} which may assume non-integer values. The  parameter  $d$ 
is independent of the   physical mechanisms that are linked to
 confinement effects in TMDCs, which simplifies   the 
 evaluation of electro-optical properties in low dimensional systems \cite{tanguy1996complex,tanguy1997analytical}. The theoretical predictions using FDSA  yields good agreement with 
experimental findings  \cite{matos99,christo93,mathieu,thilagam1997exciton}, and 
 provides qualitative insights that  could be useful in the fabrication of devices.
The  FDSA  enables   understanding of the 
underlying   quantum dynamical processes that control device operation,
and  provides valuable  information on  the
 cost effective fabrication of optical devices.

In this study, we compute the complex dielectric function of
low dimensional transition metal dichalcogenides using
a model of the exciton as a quasi-particle with arbitrary  dimensions 
 \cite{tanguy1996complex,tanguy1997analytical}.
In TMDCs, the exciton dimensionalities  are known to vary between 1.7 and 2.5 \cite{thiljap}
and may be quantified  either by the  ratio of the monolayer 
height to the exciton Bohr radius, or by the
degree of confinement  of the exciton within the monolayer plane.
To this end, we   analyze the  contributions of the  $A$ and $B$ excitons
to the broadened complex dielectric constant based on the Kramers-Kronig relations.
The  calculations linked to these relations are simplified as 
the fractional dimensionalities  
$d_A$ and  $d_A$  incorporate the blue-shifted 
 absorption shifts and broadening effects  arising
from the  quantum correlated dynamics between the $A$ and $B$ excitons
 \cite{sim2013exciton}. 
A detailed analysis of the repulsive inter-excitonic interactions 
is beyond the scope of this study, however we aim to extract
approximate estimates of the broadening effects of the $A$ and $B$ excitons
using the FDSA formalism in this work.

\section{The dielectric constant of low-dimensional excitons}

\subsection{The dielectric constant  in $d$ dimensions}

The complex dielectric function  $\epsilon_{\alpha}(E)$
that is applicable in $d$ dimensions reads as
\bea
\label{diect1}
\epsilon_{d}(E) &=& \frac{A_d \; R^{d/2-1}}{(E + \mathrm{i} \gamma)^2} \left[g_d \left(\xi(E+\mathrm{i} \gamma_b)\right)+
g_d \left(\xi(-E - \mathrm{i} \gamma_b) \right)- 2  g_d \left( \xi(0) \right) \right], \\ \label{diect2}
g_d (x) &=& \frac{2 \pi \Gamma(\frac{d-1}{2}+x)}{\Gamma(\frac{d-1}{2})^2 \; \Gamma(1-\frac{d-1}{2}+x) \; x^{d-2}}
\; \left[ \cot \pi\left(\frac{d-1}{2}-x \right)-\cot \pi(d-1) \right], \\ \label{diect3}
\xi(x) &=& \left( \frac{R}{E_g-x} \right)^{\frac{1}{2}}
\eea
where $\gamma$ is the finite width of the Lorentzian broadened
transitions and $E_g$ is the effective band gap of the bulk material that 
incorporates confinement effects.The symbol $\Gamma(x)$ denotes the 
 Euler gamma function,  $R$ is the effective Rydberg and $A_d$ parametrizes the 
exciton oscillator strength. All bound and unbound states arising from  Coulomb interactions 
are taken into account in Eq.\ref{diect1}-\ref{diect3}.
Using Eq.\ref{diect1}, we express the total dielectric function for the monolayer material
via the Sellmeier equation and 
summing the contributions  from the $A$ and $B$ excitons as
\be
\label{theorf}
\epsilon_{d}^T(E) = C + \frac{a}{b-E^2}+\epsilon_{d_A}^A(E)+\epsilon_{d_B}^B(E+\delta)
\ee
where $\epsilon_{d}^A(E)$ and $\epsilon_{d}^B(E+\delta)$ are the  dielectric function
contributions due to the $A$ and $B$ excitons respectively. The term $\delta$ quantifies the
separation between the  $A$ and $B$ excitonic peaks.  The  dimensions associated
with the $A$ ($B$) exciton is denoted by $d_A$ ($d_B$).
The finite width of the Lorentzian broadened
transitions associated with the $A$ ($B$) exciton is denoted by $\gamma_a$ ($\gamma_b$).

\begin{figure}[htp]
  \begin{center}
\subfigure{\label{Ka}\includegraphics[width=7.5cm]{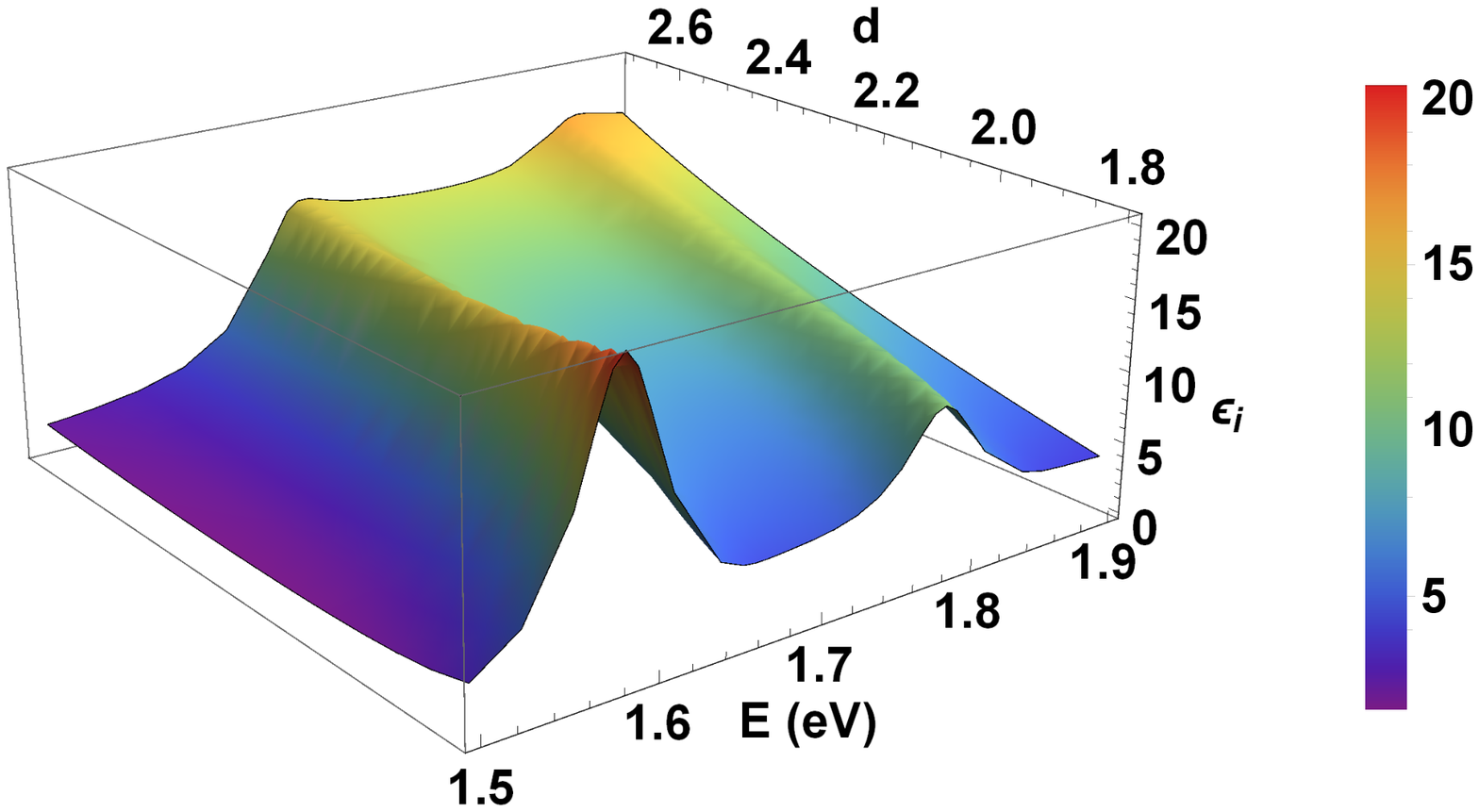}}\vspace{-1.1mm} \hspace{5.2mm}
\subfigure{\label{Kb}\includegraphics[width=7.5cm]{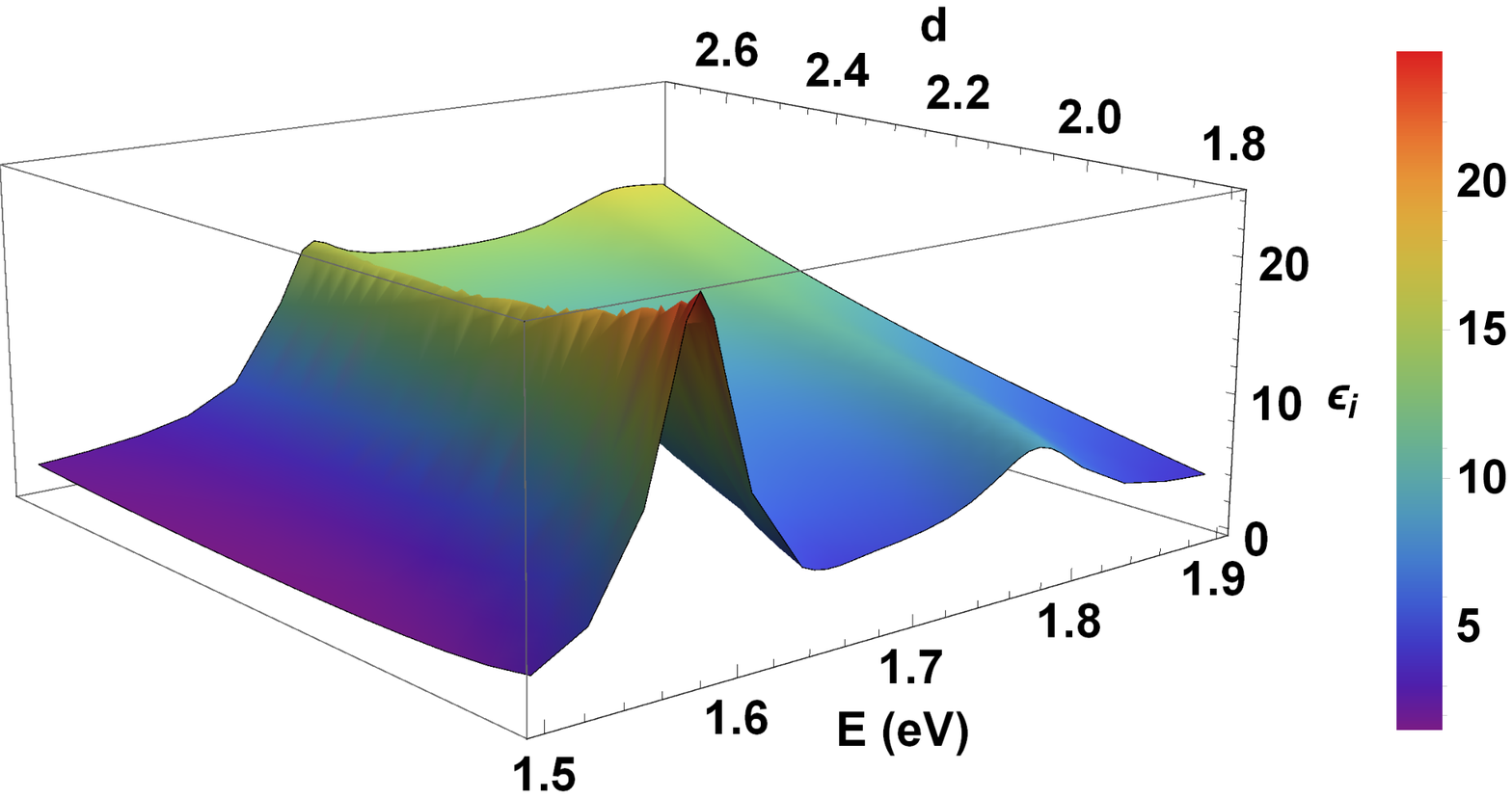}}\vspace{-1.1mm} \hspace{1mm} 
 \end{center}
     \caption{(a) Imaginary component of the dielectric
constant ($\epsilon_i$) using Eq.\ref{theorf} as a function of energy $E$ and 
the equal exciton dimensions $d_A$ = $d_B$ = $d$. We set the Lorentzian widths
$\gamma_a$ = 0.03 eV, $\gamma_b$ = 0.03 eV, the effective Rydberg $R$ = 0.015 eV
and take  $A_d \; R^{d_A/2-1}$ = 8,
$A_d \; R^{d_B/2-1}$= 4. We also set the effective band gap $E_g$ = 1.68 eV with $\delta$ = 0.2 eV.
\\
 (b) Imaginary component of the dielectric
constant using Eq.\ref{theorf} as a function of energy $E$ and 
the equal exciton dimensionalities $d_A$ = $d_B$ = $d$.
All other parameters used are the same as specified in (a) with the exception of 
$\gamma_a$ = 0.025 eV, $\gamma_b$ = 0.04 eV.}
 \label{diel}
\end{figure}

Fig.\ref{diel} a, b illustrates the decrease of  the 
imaginary component  ($\epsilon_i$) of the $A$ exciton  and  corresponding increase  of 
$\epsilon_i$ of the $B$ exciton 
with increase of  the effective dimensions of both excitons ($A$ and $B$).
A decrease in the Lorentzian widths
$\gamma_a$ and $\gamma_b$ results in  sharper excitonic peaks as is to be expected.
All  other parameters used to generate Fig.\ref{diel} a, b are listed below the figures.
The strongly confined ground exciton state  ($d \approx$ 2) contributes dominantly to the dielectric constant properties
of the anisotropic material system as seen in  Eqs.\ref{diect1} and \ref{theorf}. The contribution from  the higher bound excitons states is significantly less  as the oscillator
strengths of the transitions ($2s$, $2p$)  are 
substantially decreased in higher order exciton states.

\begin{figure}[htp]
  \begin{center}
\subfigure{\label{Ka1}\includegraphics[width=9.5cm]{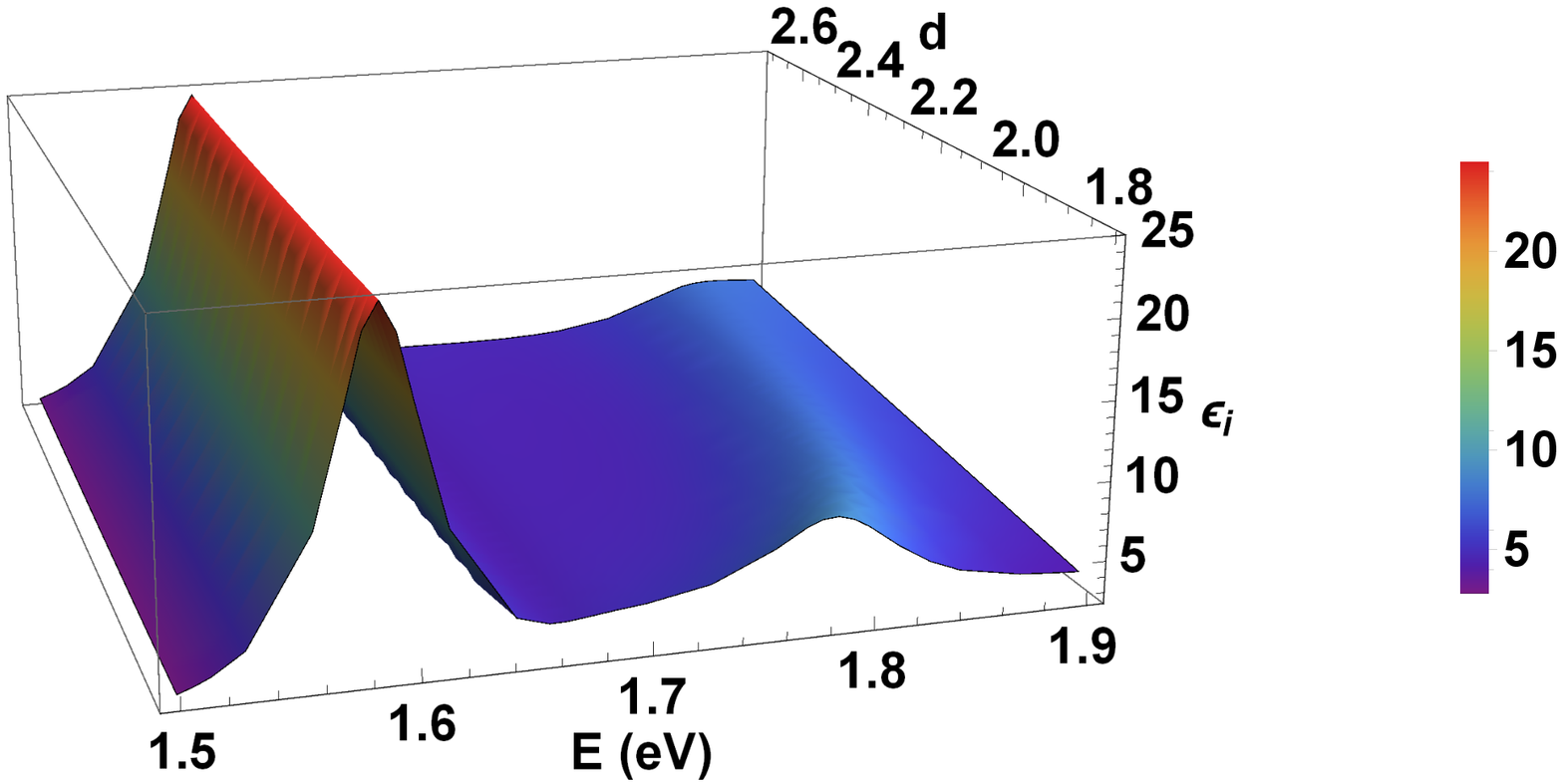}}\vspace{-1.1mm} \hspace{6.2mm}
\subfigure{\label{Kb1}\includegraphics[width=7.5cm]{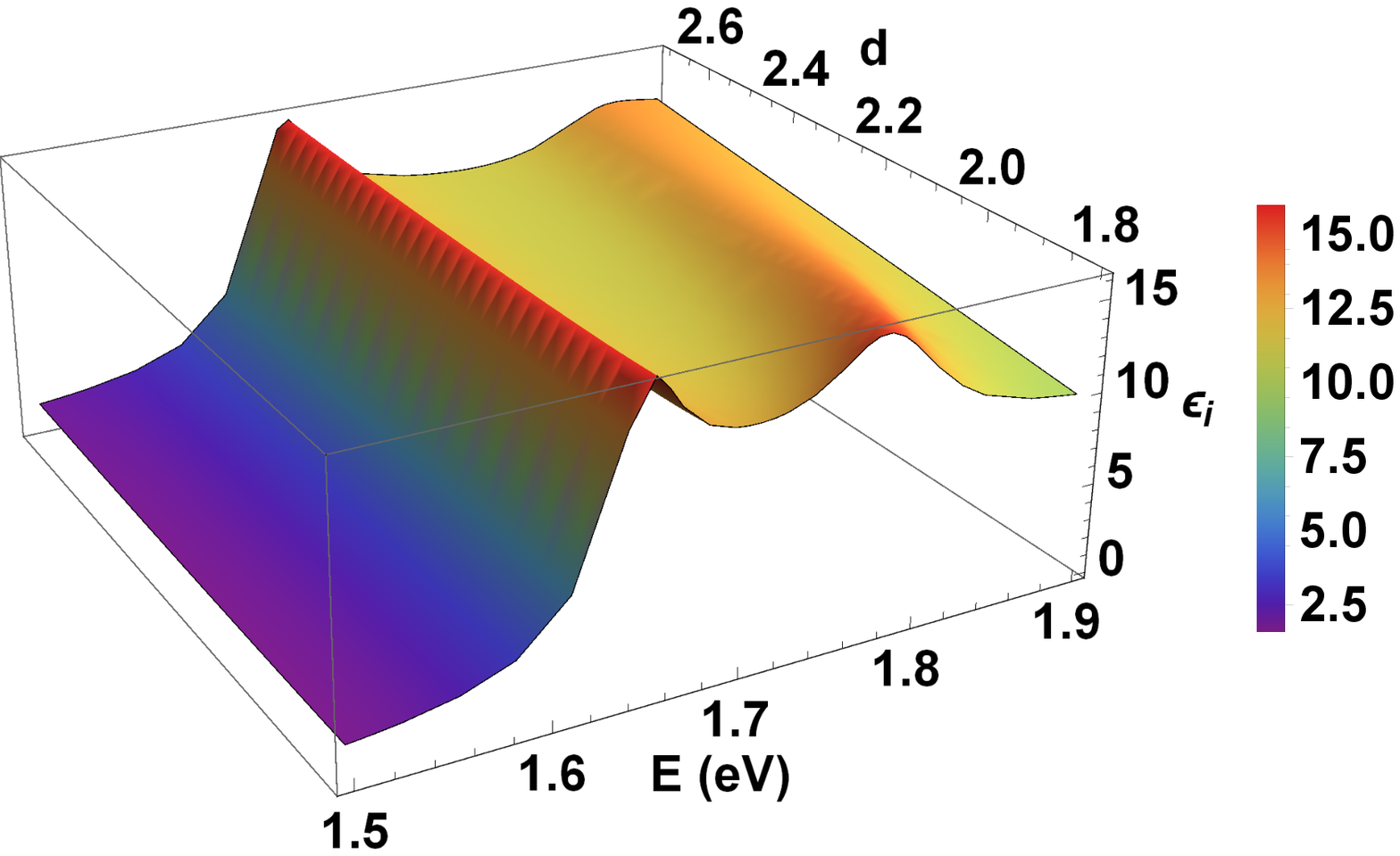}}\vspace{-1.1mm} \hspace{1mm} 
 \end{center}
     \caption{(a) Imaginary component of the dielectric
constant using Eq. \ref{theorf} as a function of energy $E$ and 
the exciton dimensions $d_B$ = d with fixed  $d_A$ = 1.8. We set the Lorentzian widths
$\gamma_a$ = 0.025 eV, $\gamma_b$ = 0.04 eV, and fix  $A_d \; R^{d_A/2-1}$= 8,
$A_d \; R^{d_B/2-1}$= 4,  the effective band gap $E_g$ = 1.68 eV and $\delta$ = 0.2 eV.
\\
 (b) Imaginary component of the dielectric
constant using Eq.\ref{theorf} as a function of energy $E$ and 
the  exciton dimensions $d_B$ = d with fixed  $d_A$ = 2.5.
All other parameters used are the same as specified in (a)  above.}
 \label{broad}
\end{figure}
Fig. \ref{broad} a, b highlights the links between the dimensionality of
the $A$ exciton,  the dielectric constant as well as
dimensionality of the $B$ exciton. The dimension of the $A$  exciton 
is fixed at  $d_A$ = 1.8 in Fig.\ref{broad} a, and at the higher
$d_A$ = 2.5 in Fig.\ref{broad} b.
 A  gradual decrease in the imaginary component  ($\epsilon_i$) of the $B$ exciton occurs
 with increase of  its effective
dimensionality. The results show that a lower exciton dimension $d_A$ = 1.8 is associated
with a weakened contribution to the dielectric constant estimates by the 
  $B$  exciton. There is partial qualitative agreement of these  results 
with experimental observations \cite{sim2013exciton} 
of correlated interactions between the $A$ and $B$ excitons which are
closely linked in momentum and energy space.

The complex spectral optical properties of low dimensional material systems
are  influenced by collision-induced excitonic linewidth broadening effects.
The non-local quantum interaction  between the $A$ and $B$ excitons is expected to 
influence  the optical properties of highly confined material systems.
A previous work has shown that the  broadened $B$ exciton linewidth is linked to a 
diminished peak spectral amplitude of the $A$ exciton \cite{sim2013exciton}. 
To this end, there is possibility that collision-induced excitonic linewidth broadening effects occurring
at one exciton ($A$ or $B$) may influence the spectral amplitude
of the adjacent exciton.  The incorporation of non-local quantum interactions
between the $A$ and $B$ excitons as carried out in an earlier work\cite{thilagam2010zeno} is expected to introduce
greater accuracy in the analysis of the mutually driven quantum correlated
interactions between the $A$ and  $B$  excitons.

\subsection{The $A$ and $B$ excitonic peaks in monolayer MoS$_2$}

The dielectric function  of the monolayer
MoS$_2$ displays three peaks  that range  from low
to high energies and are labelled as $A, B, C$  \cite{komsa2012effects}. 
The  $A$ and $B$ excitonic peaks  arise
from the  electron-hole interaction between the spin-orbit
split valence bands and the lowest conduction band at the $K$ and $K'$ points.
The $A$ ($B$) exciton is formed from the spin-up (spin-down) electrons in the conduction
band (K-point of the BZ) and spin-down (spin-up) holes in the valence bands.
The $A$ and $B$ excitons in general have almost similar behaviors,
with any  difference expected to arise from the position of holes 
in  separate valence bands. The $C$ peak is linked to electron-hole interaction
between the valence band and  the conduction band in the vicinity of
the $\Sigma$ and $\Lambda$ points \cite{qiu2013}.
Ellipsometry optical techniques which allow the precise extraction of the dielectric function
reveal the presence of the $A$ and $B$ exciton in the monolayer MoS$_2$ on SiO$_2$/Si substrates
 at 1.88 eV and 2.02 eV respectively \cite{li2014broadband}. 
There exist  three contributing factors to the dielectric constant in this region: 1) the lowest bound exciton, 2)
all other  higher bound exciton states, and 3) continuum contribution
that incorporates a Sommerfeld factor due to  Coulomb attraction. The weak higher bound exciton states 
tend to merge with increase in number of states, hence the contribution from the lowest bound exciton
is considered separately from the rest of the higher bound states.
With  increase in the exciton binding energies at lower dimensions, 
the  resonance energies of  the monolayer dielectric function are shifted from the
corresponding energies in the bulk material \cite{li2014measurement}.

\subsection{Comparison of theoretical results with experimental data: monolayer  MoS$_2$}

The   complex in-plane
dielectric functions of four monolayer TMDCs (MoS$_2$,
MoSe$_2$, WS$_2$, and WSe$_2$) \cite{li2014measurement} have been derived using
Kramers-Kronig constrained analysis
of the reflectance spectra  of the monolayers placed on fused silica substrates.
Here  we  focus on the optical
spectrum region (1.5 - 3 eV) \cite{li2014measurement} in the vicinity of the $A$ and $B$ excitonic peaks for
the Molybdenum based   monolayer materials: MoS$_2$ and MoSe$_2$.
We determine the dimensions of the $A$ and $B$ excitons ($d_A$ and $d_B$) 
as best-fit parameters based on    the experimental data of Li et. al. \cite{li2014measurement}
and the {\it NonlinearModelFit} option in the Mathematica  package.
The nonlinear model for this procedure is constructed 
using the fractional  dimensional space dielectric model of  Eq. \ref{theorf}. 
Other than the dimensions  $d_A$ and $d_B$, we also determine
$A_1$ and $A_2$ which are the respective amplitudes  the $A$ and $B$ exciton, and
the finite widths of excitonic transitions $\gamma_a$ and $\gamma_b$.
The  band gap $E_g$  and  $\delta$ which is the
separation between the  $A$ and $B$ excitonic peaks are taken 
as free parameters.

\begin{figure}[htp]
  \begin{center}
\subfigure{\label{Ka2}\includegraphics[width=8.5cm]{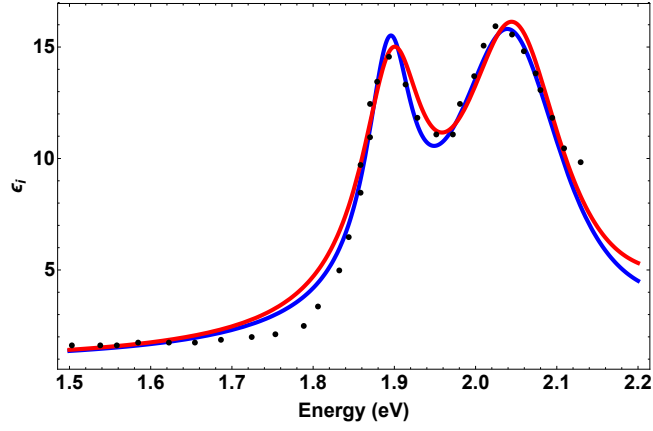}}\vspace{-1.1mm} \hspace{5.2mm}
 \end{center}
     \caption{Experimental results \cite{li2014measurement}  of the
imaginary component of the dielectric constant  of  the monolayer MoS$_2$ (dotted)
fitted using the theoretical model ($\epsilon_i$ in Eq.\ref{theorf}).
The best-fit parameters are determined via the 
 {\it NonlinearModelFit} function of Mathematica, with  the  effective Rydberg $R$ 
fixed at a specific value.
The fitted parameters for the blue line are as: $d_A$ = 2.0,  
$d_B$ = 1.95,  effective band gap $E_g$ = 2.11 eV, $\delta$ = 0.17 eV,
amplitudes $A_1$ = 2.27 and $A_2$ = 8.65  with  $R$ fixed at 55 meV.
The Lorentzian broadening widths of $\gamma_a$ = 34 meV ($A$ exciton) 
and $\gamma_b$ = 83 meV ($B$ exciton). \\ 
The   red line is obtained using $R$= 45 meV and the fitted parameters
are:  the exciton dimensions $d_A$ = 2.0 and $d_B$ = 1.95,
 effective band gap $E_g$ = 2.08 eV, $\delta$ = 0.17 eV,
amplitudes $A_1$ = 3.78 and $A_2$ = 9.37 (red line). 
The Lorentzian broadening widths of $\gamma_a$ = 45 meV ($A$ exciton) 
and $\gamma_b$ = 77 meV ($B$ exciton).
}
 \label{mos}
\end{figure}

Fig. \ref{mos} illustrates the fitting of the experimental results of the
imaginary component of the dielectric constant  \cite{li2014measurement}
for the monolayer MoS$_2$ material using the fractional  dimensional dielectric model of  Eq. \ref{theorf}. 
The best-fit parameters are extracted using the 
 {\it NonlinearModelFit} function of Mathematica, with  the  effective Rydberg $R$ 
fixed at a specific value.
A range of values (40 to 60 meV) of  $R$ for the 
bulk MoS$_2$ \cite{saigal2016exciton} has been reported, and we thus fix
$R$ at two possible values of 45 meV and 55 meV.
Using $R$= 55 meV, results of the  fitting procedure give  $d_A$ = 2.0,  
$d_B$ = 1.95,  effective band gap $E_g$ = 2.11 eV, $\delta$ = 0.17 eV,
amplitudes $A_1$ = 2.27 and $A_2$ = 8.65 (blue line). 
The Lorentzian broadening widths are obtained as  $\gamma_a$ = 34 meV ($A$ exciton) 
and $\gamma_b$ = 83 meV ($B$ exciton).
Using $R$= 45 meV, results of the  fitting procedure give  $d_A$ = 2.0,  
$d_B$ = 1.95, 
 effective band gap $E_g$ = 2.08 eV, $\delta$ = 0.17 eV,
amplitudes $A_1$ = 3.78 and $A_2$ = 9.37 (red line). 
The Lorentzian broadening widths using $R$= 45 meV are obtained as $\gamma_a$ = 45 meV ($A$ exciton) 
and $\gamma_b$ = 77 meV ($B$ exciton). 
The estimates for  band gap $E_g$  and  the energy difference $\delta$ between $A$ and $B$  excitonic peaks are 
in reasonable agreement with those obtained in an earlier work \cite{rama,komsa2012effects}.
The  results here confirm that  the ground exciton $A$ and $B$ states ($d \approx$ 2) contribute
dominantly to the dielectric constant properties of the monolayer  MoS$_2$ in the optical region (1.5 - 3 eV), with the exciton  dimensionality playing a critical role
in determining the dielectric properties
of monolayer systems.  

Using the hydrogenic binding energy relation 
\be
\label{bindH}
E_{b} = {\frac{R}{\left (1+\frac{d-3}{2} \right )^2}}
\ee
we estimate the binding energy of the $A$  exciton to be about  220 meV (using $R$= 55 meV)
and 180 meV ($R$ = 45 meV). The  $B$  exciton
has binding energy of about  244 meV ($R$ = 55 meV)
and 200 meV ($R$ = 45 meV),  due to the higher hole mass
of one of  the  spin-orbit split valence bands. 
These predicted results using Eq. \ref{bindH} are  substantially smaller than the 
binding energies ($\approx$ 0.85 eV) obtained in  earlier works \cite{chei12,rama,komsa2012effects}.
but  in fair agreement with the
well-converged first principle Bethe-Salpeter derived estimates (200 meV to 300 meV)
by A. Molina-Sanchez et al. \cite{molinaspin}, and also
with those of Bergh{\"a}user et al \cite{berghauser2014analytical}
who obtained binding energies of 455 (465 meV) for 
$A$ ($B$) excitons respectively The scaling relation between band gap and exciton binding energy 
of 2D systems \cite{jiang2017scaling} implies typical exciton binding energies of around 400 to 500 meV for the monolayer MoS$_2$ on a silicon substrate. There is also some consistency of our results with
the exciton binding energy of 0.3 eV  computed using $R$= 77 meV  in Ref.\cite{soh2018optical},
and also with  the photoluminescence excitation spectroscopy results of  monolayer MoS$_2$ on fused silica
 which provided an exciton binding energy of 0.44 eV \cite{hill2015observation}.
In the case of  the  MoS$_2$ monolayer in vacuum\cite{berghauser2014analytical},
larger binding binding energies of 860 and 870 meV were obtained for the $A$, $B$ excitons.
Due to the screening effects induced by the substrate,  the exciton binding energy is reduced
which correlates with an increased exciton dimensionality.

For  TMDC monolayers supported on
SiO$_2$ substrates, the photoluminescence linewidth is generally larger than 10 meV at low temperatures.
Based on the fitting results, the Lorentzian broadening widths of $\gamma_a$ = 34 meV (45 meV) ($A$ exciton) and $\gamma_b$ = 83 meV (77 meV) ($B$ exciton) at  $R$= 55 meV (45 meV) are of the same order of the broadened width of  60 meV  evaluated in an earlier work \cite{molina2016temperature} for the monolayer MoS$_2$. The broadening kinetics associated with the $A$ ($B$) exciton are known to arise from several sources: exciton-optical phonon \cite{molina2016temperature,dey2016optical,thilagam2016exciton}, exciton-acoustic phonon interactions \cite{molina2016temperature,thilrelaxjap}, exciton-impurity scattering \cite{ajayi2017approaching} and,
 inter-excitonic scattering  processes  \cite{sim2013exciton,cadiz2017excitonic}.
The larger broadened width of the $B$ exciton can be attributed to its origin at the second valence band 
with availability of increased recombination pathways  \cite{molina2016temperature} compared to the $A$ exciton.
The differences in  hole energy and  population densities 
between the two distinct valence bands linked to the $A$ and $B$  excitons
also account for  broadening width differences in $\gamma_a$ and $\gamma_b$.

\subsection{Comparison of theoretical results with experimental data: monolayer  MoSe$_2$}

\begin{figure}[htp]
  \begin{center}
\subfigure{\label{Kb2}\includegraphics[width=8.5cm]{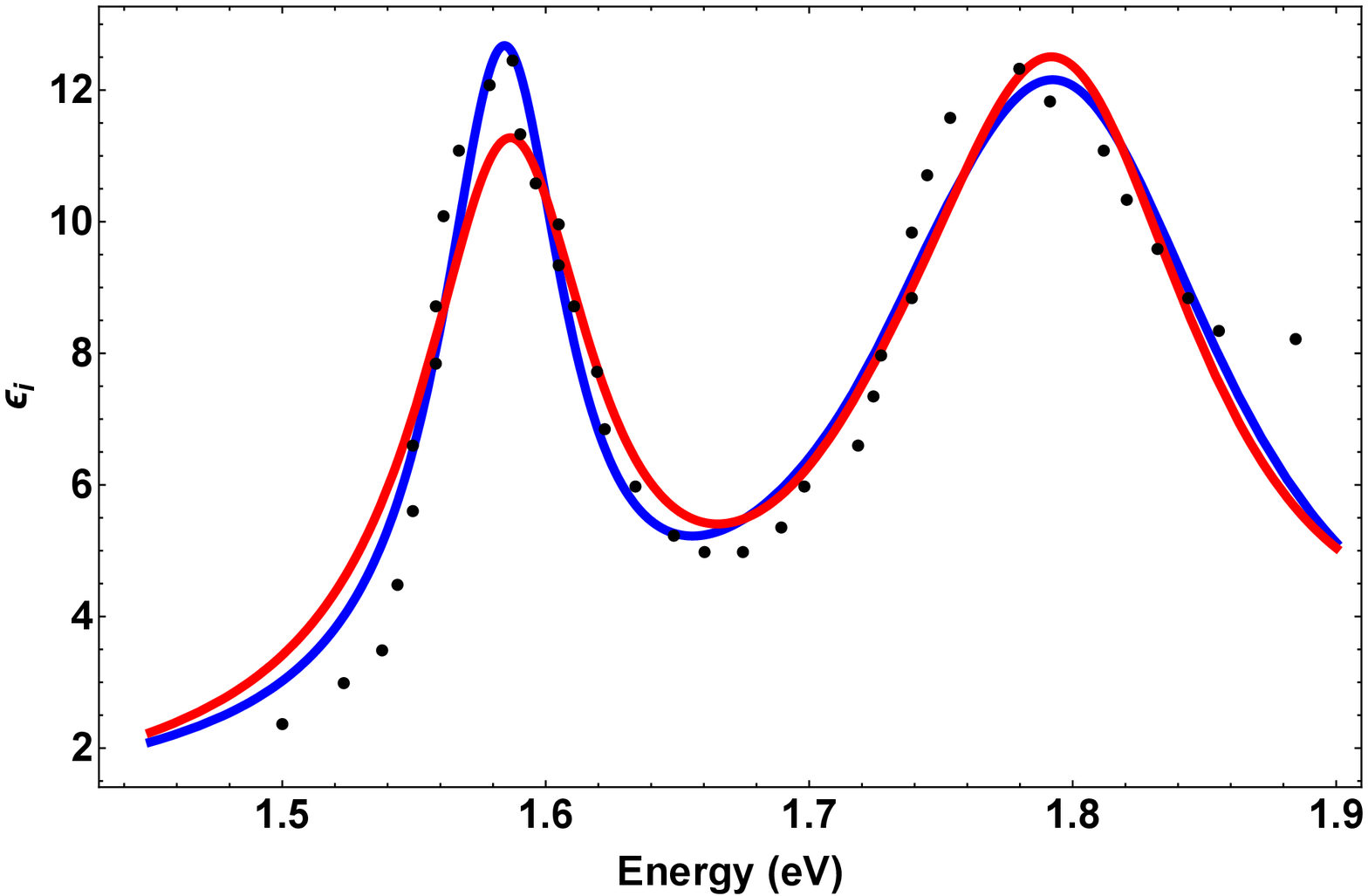}}\vspace{-1.1mm} \hspace{1mm} 
 \end{center}
   \caption{Experimental results \cite{li2014measurement}  of the
imaginary component of the dielectric constant  of  the monolayer MoSe$_2$ (dotted)
fitted using the theoretical model ($\epsilon_i$ in Eq.\ref{theorf}).
The best-fit parameters derived using the 
 {\it NonlinearModelFit} model ( blue line) are as follows: $d_A$ = 2.05,  
$d_B$ = 1.98,  effective band gap $E_g$ = 1.77 eV, $\delta$ = 0.24 eV,
amplitudes $A_1$ = 1.50 and $A_2$ = 5.44  with  $R$ fixed at 50 meV.
The Lorentzian broadening widths of $\gamma_a$ = 28 meV ($A$ exciton) 
and $\gamma_b$ = 82 meV ($B$ exciton). \\ 
The   red line is obtained using $R$= 40 meV and the fitted parameters
are:  the exciton dimensions $d_A$ = 2.0 and $d_B$ = 1.98,
 effective band gap $E_g$ = 1.74 eV, $\delta$ = 0.22 eV,
amplitudes $A_1$ = 2.30 and $A_2$ = 5.72 (red line). 
The Lorentzian broadening widths of $\gamma_a$ = 40 meV ($A$ exciton) d
and $\gamma_b$ = 70 meV ($B$ exciton).
}
 \label{mose}
\end{figure}

In few-layer MoSe$_2$,  the
indirect bandgap and direct bandgap are nearly degenerate unlike the MoS$_2$ system \cite{tongay2012thermally}.
There exist subtle differences between the monolayers MoSe$_2$ and MoS$_2$
 in terms of the reduced exciton mass, dielectric constants  \cite{kylanpaa2015binding} and 
band gaps \cite{tongay2012thermally}. 
Fig. \ref{mose} illustrates the fitting of the experimental results of the
imaginary component of the dielectric constant  \cite{li2014measurement}
for the monolayer MoSe$_2$ with the fractional  dimensional dielectric model of  Eq. \ref{theorf}. 
By fixing the effective Rydberg $R$ at 50 meV,
we get from the  fitting procedure:  $d_A$ = 2.05,  $d_B$ = 1.98,
 effective band gap $E_g$ = 1.77 eV, $\delta$ = 0.24 eV,
amplitudes $A_1$ = 1.5 and $A_2$ = 5.4  and 
the Lorentzian broadening widths of $\gamma_a$ = 29 meV ($A$ exciton) and $\gamma_b$ = 82 meV ($B$ exciton).
Using Eq. \ref{bindH} and the exciton dimensions obtained here, the binding energy of the $A$ exciton is evaluated as 181 meV and that of the $B$ exciton as 208 meV.
With the effective Rydberg $R$ at 40 meV,
we get the following results from the  fitting procedure:  $d_A$ = 2.01,  $d_B$ = 1.98,
 effective band gap $E_g$ = 1.74 eV, $\delta$ = 0.22 eV,
amplitudes $A_1$ = 2.3 and $A_2$ = 5.7  and 
the Lorentzian broadening widths of $\gamma_a$ = 40 meV ($A$ exciton) and $\gamma_b$ = 70 meV ($B$ exciton).
Using Eq. \ref{bindH} and the exciton dimensions obtained for
$R$ = 40 meV, the binding energy of the $A$ exciton is evaluated as 160 meV and that of the $B$ exciton as 167 meV. The  Lorentzian broadening widths of the $B$ exciton is larger than the $A$ exciton for reasons mentioned in the earlier section for the monolayer  MoS$_2$.

\subsection{Comparison of theoretical results with experimental data: monolayer  WSe$_2$}

The exciton binding energy in 
monolayers of tungsten diselenide (WSe$_2$) has been  determined via optical techniques to be 0.37 eV
with a band gap energy of 2.02eV . Due to the strong spin-orbit coupling in WSe$_2$,
the energy separation between the $A$ and $B$ exciton is large (about 0.43 eV) \cite{he2014tightly}.
Another study \cite{hanbicki2015measurement} revealed a  much higher  experimentally determined exciton binding energy of 0.79 eV showing the large variations in binding energies that exists amongst different experimental and theoretical
groups. Experimental
determination of the exciton binding energy of monolayer WSe$_2$ 
was noted to be 240 meV on sapphire substrate while on gold the exciton binding 
 decreased to 140 meV \cite{park2018direct}. As expected, 
the enhanced screening by the metal substrate results in lower  binding
and a larger dimensionality 
for the exciton.

\begin{figure}[htp]
  \begin{center}
\subfigure{\label{Kb3}\includegraphics[width=8.5cm]{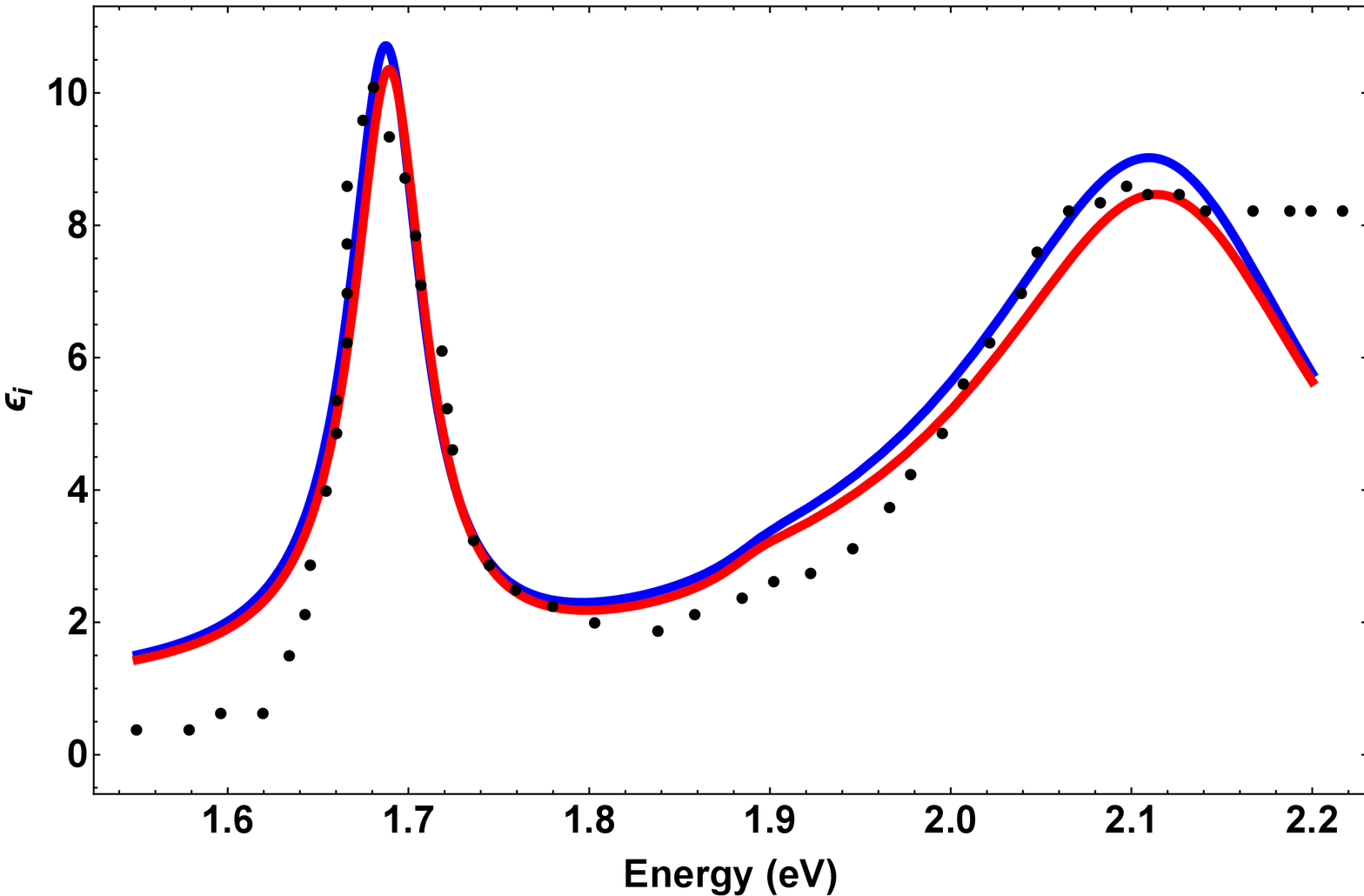}}\vspace{-1.1mm} \hspace{1mm} 
 \end{center}
   \caption{Experimental results \cite{li2014measurement}  of the
imaginary component of the dielectric constant  of  the monolayer WSe$_2$ (dotted)
fitted using the theoretical model ($\epsilon_i$ in Eq.\ref{theorf}).
The best-fit parameters derived using the 
 {\it NonlinearModelFit} model (see blue line) are as follows: $d_A$ = 1.85,  
$d_B$ = 1.82,  effective band gap $E_g$ = 1.92 eV, $\delta$ = 0.45 eV,
amplitudes $A_1$ = 1.3 and $A_2$ = 8.51, and the Rydberg    $R$  is obtained as 42 meV.
The Lorentzian broadening widths of $\gamma_a$ = 24 meV ($A$ exciton) 
and $\gamma_b$ = 116 meV ($B$ exciton). \\ 
The   red line is obtained using  parameters derived from the fitting procedure: $R$= 52 meV , the exciton dimensions $d_A$ = 1.95 and $d_B$ = 1.92,
 effective band gap $E_g$ = 1.91 eV, $\delta$ = 0.45 eV,
amplitudes $A_1$ = 1.12 and $A_2$ = 7.01 (red line). 
The Lorentzian broadening widths of $\gamma_a$ = 24 meV ($A$ exciton) 
and $\gamma_b$ = 117 meV ($B$ exciton).
}
 \label{Wse}
\end{figure}

Fig. \ref{Wse} illustrates the fitting of the experimental results of the
imaginary component of the dielectric constant  \cite{li2014measurement}
for the monolayer WSe$_2$ with the fractional  dimensional dielectric model (see  Eq. \ref{theorf}). Instead of  using fixed estimates for the effective  Rydberg $R$,  we allow
the {\it NonlinearModelFit} function to yield appropriate values for $R$.
 We obtain two possible estimates: $R$= 42 meV and 52 meV and in both cases,
we obtain slightly higher value for the dimensionality of the $A$ exciton
compared to the $B$ exciton. The binding energies of the  $A$ ($B$) exciton
is about 231 meV (246 meV) for both values of the effective  Rydberg.
These binding energy estimates are comparable to that (240 meV) obtained on the insulator
sapphire substrate by Part et. al. \cite{park2018direct}.
The electronic band gap of 1.89 eV for the monolayer WSe$_2$/sapphire substrate
configuration \cite{park2018direct} agrees well with the estimate of
 $E_g$ = 1.91 eV obtained using the  {\it NonlinearModelFit} model in 
Fig. \ref{Wse}. 
 
The large energy separation between the $A$ and $B$
exciton state of $\delta$ = 0.45 eV derived here is consistent with an earlier result
(0.43 eV) \cite{he2014tightly}. We note that the Lorentzian broadening widths of $\gamma_b$ = 117 meV ($B$ exciton) is higher than the corresponding widths for the monolayer 
MoS$_2$ and MoSe$_2$ (see Table \ref{tabcom}). The large broadened width of the WSe$_2$  can be attributed to the enhanced 
recombination pathways at the location of the $B$ exciton due to its  large energy separation  from the $A$ exciton. It is likely that 
differences in  hole  population densities and exciton-phonon interactions between the two
excitons further contribute to the wide variations seen in their Lorentzian broadening widths.
Nevertheless  further quantitative analysis  is needed
to identify the underlying factors that give rise to the  wide difference in broadening widths
between the $A$ and $B$ excitons.
The theoretical fit around the region of the $B$ exciton as shown in Fig. \ref{Wse} also
indicates that further refinement is needed for the fractional dimensional model used in this study. This will be considered in future works.

Table \ref{tabcom} shows a comparison of the various best-fit parameters using
imaginary component of the dielectric constant  of  the monolayer MoS$_2$,
 MoSe$_2$ and  WSe$_2$. The binding energies of the MoSe$_2$ based excitons are lower than
the binding energies of the monolayer MoS$_2$, in agreement with 
earlier results \cite{berkelbach2013theory,komsa2012effects}.
The energy difference between the A and B transitions
in the monolayer MoSe$_2$ is larger than that in MoS$_2$,  consistent with the result
of Li et al \cite{li2014measurement} and Liu et al \cite{liu2013three}.
In comparison to the Molybdenum based monolayers,  WSe$_2$ displays
a larger Lorentzian broadening width, $\gamma_b$. 

\section{Conclusion}
In summary, we have examined the dielectric properties of low dimensional transition metal
dichalcogenides  using  a fractional  dimensional space model of the  complex dielectric
constant  expression. Such a  model 
simplifies  the analysis 
 of electro-optical properties in monolayer systems,
and enables easy comparison  between the different monolayer  TMDCs.
Our results show that the 
 ground exciton state  ($d \approx$ 2) contributes 
strongly to the dielectric constant properties
of the material system.  For the monolayer materials (MoS$_2$, MoSe$_2$ and WSe$_2$) examined here the oscillator strengths of the higher order exciton state
 transitions   are substantially suppressed. 
For the purpose of  rationalizing the excitonic 
features of different TMDCs, we extract
the dimensionalities of the $A$ and $B$ excitons as well
the broadening  widths of excitonic transitions $\gamma_a$ and $\gamma_b$
using the theoretical model here and the
experimental data of Li et. al. \cite{li2014measurement}
based on  monolayers (MoS$_2$,
MoSe$_2$, WSe$_2$) placed on fused silica substrates.
We also extract the  band gap $E_g$   and the
separation between the  $A$ and $B$ excitonic peaks $\delta$ using the 
fractional  dimensional space model and a
 {\it NonlinearModelFit} function, with  the  effective Rydberg $R$ 
fixed at a specific value in some cases.
 
 There is  good agreement between our theoretical predictions of the  A-B splitting
and the experimental results \cite{li2014measurement} for the monolayer TMDCs (MoS$_2$,
MoSe$_2$,  and WSe$_2$). The exciton binding energies agree reasonably well with exciton  binding energy estimates obtained in earlier works \cite{molinaspin,berghauser2014analytical,soh2018optical,
hill2015observation,park2018direct}. The results in this study
show that the $B$ exciton has a marginally lower dimensionlity
than the $A$ exciton in all the examined monolayer TMDCs. 
Moreover the   non integer-dimensional occurrences of excitonic
dimensionality  presents as a reliable feature in  computational 
modeling.  The broadening width estimates of the $A$ and $B$ excitons ($\gamma_a$, $\gamma_b$)
derived using the fractional dimensional model reveal a larger 
width for the $B$ exciton which arises
from differences in the environment of the two  valence bands.
The screening effects of the substrate influences the 
exciton dimensionality, and its binding energy. For instance, the 
the enhanced screening by a metal substrate results in a larger exciton dimensionality 
 compared to a monolayer placed on an insulator substrate.
Thus exciton dimensionalities can be controlled via the dielectric
environment presented by the substrate. 

The results in this study show that a lower  $A$ exciton dimension is associated
with a weakened contribution to the dielectric constant by the 
  $B$  exciton. This is in  qualitative agreement  
with experimental observations \cite{sim2013exciton} 
which show strong correlated inter-excitonic
dynamics between the $A$ and $B$ excitons which are
closely linked in momentum and energy space. 
Further understanding of the origin of the correlated dynamics between the $A$ and $B$ excitons
and associated blue-shifted excitonic absorption
could be useful for the design of quantum coupled optical devices.
Lastly the fractional dimensional model of the complex dielectric function
is useful in interpreting experimental data and for
 making predictions for properties  of monolayer systems
that  are not accessible via current experimental techniques.

\eject

\begin{table*}
    \caption{\label{tabcom} Comparison of best-fit parameters obtained
using the {\it NonlinearModelFit} function and the fractional  dimensional 
space model of the  complex dielectric constant 
for  monolayer MoS$_2$,  MoSe$_2$ and WSe$_2$. The effective Rydberg $R$ is fixed for MoS$_2$ and
 MoSe$_2$ while it is extracted via the fitting procedure for WSe$_2$ (estimates with superscript).
The terms  $d_A$ and $d_B$ denote dimensionalities, 
$A_1$ and $A_2$ are the respective amplitudes of the $A$ and $B$ exciton, and
the respective finite widths of excitonic transitions are denoted by $\gamma_a$ and $\gamma_b$.
The separation between the  $A$ and $B$ excitonic peaks is denoted by  $\delta$.
 \\ \\
\newline
\newline}
\begin{tabular}{|c|c|c|c|c|c|c|c|c|c|c}
        \hline
        \hline
System		 ~ &\;$R$ (meV) \; &\;$d_A$ \; & \;$d_B$ \;&
\;$\delta$ (eV) \;& \; $E_g$ (eV) \; & $\gamma_a$  (meV)& $ \; \gamma_b$  (meV) & $A_1$, $A_2$ & $A$, $B$ Exciton binding (meV)\\
        \hline
MoS$_2$  & 55 & 2.0 & 1.95 & 0.17 & 2.11 & 34 & 83 & 2.27, 8.65 & 220, 244 (cf 200 to 300 \cite{molinaspin})\\
MoS$_2$  & 45 & 2.0 & 1.95 & 0.17 & 2.08 (1.90, 2.05 \cite{makatom}) & 45 & 77 & 3.78, 9.37 & 180, 200 (455, 465  \cite{berghauser2014analytical})\\
MoSe$_2$ & 50 & 2.05 & 1.98 & 0.24 & 1.77 & 28 & 82 & 1.50, 5.44 & 181, 208  \\
MoSe$_2$  & 40 & 2.0 & 1.98 & 0.22 & 1.74 & 40 & 70 & 2.30, 5.72 & 160, 167  \\
WSe$_2$ & 42$^*$ & 1.85 & 1.82 & 0.45 & 1.92 (cf 1.89 \cite{park2018direct}) & 24 & 116 & 1.3, 8.51 & 231, 246 (cf 240 \cite{park2018direct}) \\
WSe$_2$  & 52$^*$ & 1.95 & 1.92 & 0.45 & 1.91 & 24 & 117 & 1.12, 7.01 & 231, 246  \\
        \hline
        \hline
    \end{tabular}
\end{table*}

\end{document}